\documentclass[twocolumn,prl,showpacs]{revtex4}

\usepackage{graphicx}
\usepackage{rotating}
\usepackage{amsmath}
\usepackage{amsfonts}
\usepackage{amssymb}
\usepackage{enumerate}
\usepackage{longtable}
\usepackage{dcolumn}% Align table columns on decimal point
\usepackage{bm}

\begin{document}

\newcommand{\be}{\begin{equation}}
\newcommand{\ee}{\end{equation}}
\newcommand{\bn}{\begin{eqnarray}}
\newcommand{\en}{\end{eqnarray}}

\title{A Quantum Chemistry Plus Dynamical Mean-Field Approach for Correlated
Insulators: Application to $La_{2}CuO_{4}$.}

\author{M. S. Laad, L. Hozoi, and L. Craco}
\author{}
\affiliation{Max-Planck-Institut f\"ur Physik komplexer Systeme,
%N\"othnitzer Strasse 38,
01187 Dresden, Germany}

\date{\rm\today}

\begin{abstract}

 While the traditional local-density approximation (LDA) cannot describe Mott 
insulators, {\it ab-initio} determination of the Hubbard $U$, for example, 
limits LDA-plus dynamical mean field theory (DMFT) approaches.  Here, we 
attempt to overcome these bottlenecks by achieving fusion of the quantum 
chemistry (QC) approach with DMFT.  QC+DMFT supplants the 
LDA bandstructure by its QC counterpart as an input to DMFT.  Using 
QC+DMFT, we show that undoped $La_{2}CuO_{4}$ is a $d$-Mott insulator, and 
qualitatively discuss the circulating current- and incoherent metal phase, at 
small but finite hole doping.  Very good quantitative 
agreement with experimental photoemission- and optical spectra constitutes 
strong support for efficacy of QC+DMFT.  Our work thus opens a new avenue for 
truly {\it ab-initio} correlation-based approaches to describe correlated 
electronic systems in general.

\end{abstract}
     

\maketitle

%\section{INTRODUCTION}

High-$T_{c}$ superconductivity (HTSC) in quasi-two dimensional ($2D$) cuprates
is an outstanding, unsolved problem in modern condensed matter physics.  These
materials are stochiometric Mott insulators (MI).  Upon hole doping ($x$), a 
highly
unusual, $d$-wave pseudogapped ``metal'' with strongly non-Fermi liquid (nFL) 
properties {\it smoothly} evolves into the ``strange metal'' around optimal 
doping ($x_{c}$), where singular responses reminiscent of a $1D$ Luttinger 
liquid are observed~\cite{[pwa]}.  At lower $T$, $d$-wave superconductivity, 
peaking around $x_{c}$, is seen.  Upon overdoping, $d$-SC rapidly disappears along 
with a rapid crossover to a low-$T$ FL metal.  These unique observations defy  
understanding in any FL picture, forcing one to search for non-FL alternatives.
 
In reality, cuprates are charge-transfer (CT) ``Mott'' systems. $(i)$ Strong 
particle-hole asymmetry is indeed 
necessary~\cite{[schulthess]} for a quantitative description. LDA based 
studies~\cite{[pava]} suggest that the maximum $T_{c}^{max}$  
correlates with $t'/t$.  Moreover, $t''/t'$ is controlled by 
{\it axial} orbitals.  
$(ii)$ The role of these apical orbitals in cuprates is ill-understood; they 
may be {\it relevant} for the ``hidden order'' in the pseudo-gap
(PG) phase, presumed to be of the circulating current (CC) 
type~\cite{[varma],[kapit],[circ]}. If {\it true}, how does this constrain a 
{\it minimal} model for cuprates~\cite{[pwa-book],[yin],[varma]}? 
$(iii)$ A host of experiments~\cite{[arpes],[raman]} clearly reveal the 
${\bf k}$-space differentiation of quasiparticle (QP) states in doped cuprates. 
The normal-state PG has the {\it same} $d$-wave symmetry 
as the SC at lower $T$.  Quantum oscillation measurements show that small hole
pockets for small $x$ evolve, possibly via multiple electron- and hole-like
sheets~\cite{[proust]}, into a full, Luttinger Fermi surface (FS) around 
optimal doping ($x_{opt}$). 
$(iv)$ Finally, near $x_{opt}$, 
singular low energy responses suggest a branch cut, rather than a pole-like
 analytic structure of the one-particle Green's function, $G({\bf k},\omega)$,
near $E_{F}$~\cite{[pwa]}. This goes hand-in-hand with FS reconstruction at
$x_{opt}$~\cite{[daou]}.  Are these findings linked to a possible quantum 
critical point
(QCP) at $x\simeq x_{opt}$, as in $f$-band compounds~\cite{[cerhin5]}?
If so, does the PG state have a ``hidden'' order? 
 Developing an {\it ab-initio}
formulation capable of reconciling $(iii)-(iv)$ above; i.e, the quantum
oscillation (dHvA) data and angle-resolved photoemission (ARPES) 
dispersion with one- and two particle {\it dynamical} responses is a challenge 
for theory, and has hitherto been studied within effective models~\cite{[palee],[beli]}.  {\it If}, however, the $d$-PG phase indeed carries CC
order,~\cite{[varma],[kapit],[circ]} these issues must be studied using an 
extended Hubbard model involving {\it both} planar- and apical $p-d$ states~\cite{[pava],[circ]}.
   
 In light of $(i)-(iv)$, inclusion of strong electronic correlations in a 
{\it multi-band} Hubbard model (with planar and apical $p-d$ states) is mandatory.
Here, we extend earlier work~\cite{[liviu-mukul]} to  compute the 
{\it dynamical} responses by marrying the quantum-chemical (QC) 
band structure with multi-orbital DMFT.  We ``derive'' an {\it effective} model
for cuprates that, by construction, is consistent with $(i)-(iii)$
above. We show how {\it both} one- and two-particle spectra (PES {\it and} optics)
in the CT-MI  are {\it quantitatively} described by QC+DMFT.  This is a novel 
theoretical route, hitherto unexplored, and avoids the ``$U$'' problem in 
LDA-based approaches~\cite{[imada-solovyev]}, as detailed below.  
      
   Earlier QC work is essentially
 a {\it variational} calculation, using a multi-configurational
wave-function (similar to that used for the one-band Hubbard model~\cite{[sorella]}, but including local {\it and} nearest neighbor (n.n) $p-d$ and $d-d$ 
excitations), and quantitatively captures the 
strong renormalization of a doped carrier in a MI~\cite{[liviu-mukul]}.  Short
range electronic (AF spin) correlations anisotropically renormalize the 
{\it bare} band structure, leading to ${\bf k}$-space differentiation,
as derived before in the one-band context~\cite{[imada],[kotliar]}, and in 
good {\it quantitative} agreement with {\it both}, the ARPES dispersion~\cite{[ino]}, {\it and} the Shubnikov-de Haas data~\cite{[hussey],[proust]}.  
The nodal (N) QPs have 
predominantly {\it planar} character, while the anti-nodal (AN) QPs have 
significant mixing of apical $d_{z^{2}}-p_{z}$ states.  However, the QC 
work cannot  compute {\it dynamical} excitation spectra.  To address
$(ii)$ and $(iv)$ above, the QC work needs to be ``married'' with DMFT/cluster-DMFT
calculations.  Cluster-DMFT successfully reproduces various anomalous
responses in cuprates, but within an effective model framework.  Here, we harmonize the successes of the QC method and DMFT-like approaches for the
CT-MI phase in an {\it ab-initio} framework; the doped case will be treated 
separately.

Labelling the $d_{x^{2}-y^{2}}-p_{\sigma}$ and $d_{z^{2}}-p_{z}$ bands found in the QC work by
``$1,2$'' leads to an {\it effective} ``two-band'' Hubbard model, $H=H_{0}+H_{1}+H_{mix}$, where
$H_{0}=\sum_{{\bf k},a=1,2,\sigma}\epsilon_{{\bf k},a}c_{ka\sigma}^{\dag}c_{ka\sigma} + \sum_{i,a=1,2}\Delta_{a}n_{ia\sigma}$,

\be
H_{1}=U\sum_{i,a=1,2}n_{ia\uparrow}n_{ia\downarrow} + U'\sum_{i,a\ne b}n_{ia}n_{ib} - J\sum_{i,a\ne b}{\bf S}_{ia}.{\bf S}_{ib}
\ee
and

\be
H_{mix}=\sum_{{\bf k},\sigma}t_{m}({\bf k})(c_{k1\sigma}^{\dag}c_{k2\sigma}+h.c)  
\ee
where $t_{m}({\bf k})=t_{m}$(cos$k_{x}$-cos$k_{y}$) is the $d$-wave form factor associated with 
intersite, inter-orbital $d_{x^{2}-y^{2}}-d_{z^{2}}$ one-particle hopping. ($\Delta_{1}-\Delta_{2})=1.15$~eV and $U\simeq 5.0$~eV is 
chosen from the QC result ($E_{N+1}+E_{N-1}-2E_{N}\simeq 5.0$~eV), while $J=0.6$~eV is taken from atomic tables, and $U'\simeq U-2J=3.8$~eV.    
  As for the hoppings, $t=0.45$~eV is chosen to be its ``bare'' value, while 
($t',t'',t_{m}=0.01,0.075,0.2$~eV) in $H$ are renormalized (by short-range static AF correlations) values taken from QC results~\cite{[liviu-mukul]}.  
This is because DMFT mainly 
renormalizes $t$, but the farther-neighbor hoppings are renormalized by
non-local correlations beyond DMFT; we take the QC (static) renormalizations 
for these as an input into the DMFT machinery.   
Since QC already treats the effect of {\it static} 
correlations,~\cite{[liviu-mukul]}, we do not include the 
static Hartree contribution in the DMFT treatment, thereby avoiding double 
counting of such terms.  
QC+DMFT now treats the effects of {\it local} dynamical 
correlations and non-local, {\it static} correlations on carrier dynamics in a 
single picture.  The two dispersive
bands are then given as $\epsilon_{1}({\bf k})=-2t(c_{x}+c_{y})-2t'c_{x}c_{y}-2t''(c_{2x}+c_{2y})$ and $\epsilon_{2}({\bf k})=-2t_{z}(c_{x}+c_{y})$ with
$t_{z}<<t,t',t'',t_{m}$.  Here, $c_{\alpha}=$cos$(k_{\alpha})$ with $\alpha=x,y,z$.  While similar values for the bare hoppings are found in LDA
approaches, an advantage of QC is an {\it ab-initio} estimate of
the Hubbard $U$ within a correlated formulation, avoiding the problems 
associated with LDA in this context~\cite{[imada-solovyev]}.  Also, in contrast to LDA-based approaches, the QC
 work~\cite{[liviu-mukul]} gives, remarkably, a $d$-wave form-factor, and 
strong, local, $d$-shell correlations.  Our two-band model is thus an extended 
Anderson lattice model (EALM).  Interestingly, 
Yin {\it et al.}~\cite{[yin]} derive a similar two-band-like model from LDA+$U$.  In contrast to our QC results, however, without the $d$-wave hybridization.    Within QC+DMFT, $t_{m}(k)$ gives a $d$-CT-Mott insulator, as we show below.  

\begin{figure}[thb]
\includegraphics[width=\columnwidth]{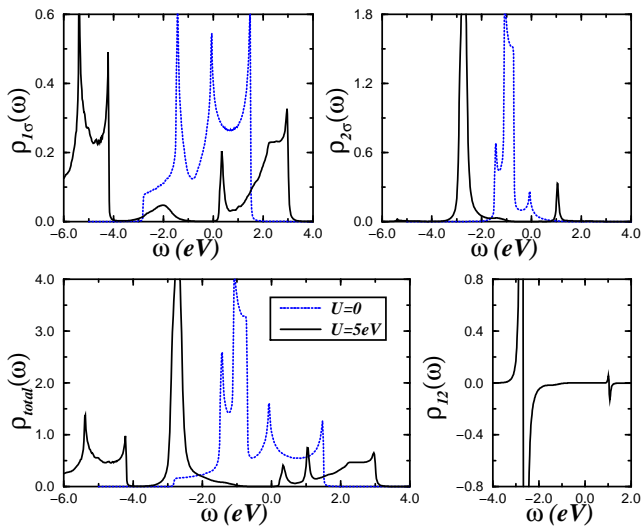}
\caption{The ``orbital resolved'' and the total many-particle density of
states (DOS) for the two-orbital model at half-filling, $\langle n\rangle=3$.  
Notice the appearance of ``orbital'' dependent Mott-Hubbard gaps, and the 
$d$-wave insulator, as explained in the text.}
\label{fig1}
\end{figure}

We solve $H$ within DMFT using the multi-orbital iterated perturbation theory
(MO-IPT) as the impurity solver~\cite{[laadv2o3]}.  Though not numerically 
``exact'', it has many advantages: (i) it gives quantitatively accurate results
for band-fillings upto half-filling~\cite{[laadv2o3]} at arbitrary 
$T$, (ii) self-energies can be easily extracted, and (iii) it is numerically 
very effcient.  More ``exact'' solvers either cannot reach low $T$ of
interest (QMC)~\cite{[liebsch]} or are prohibitively costly in {\it real},
multi-orbital cases (NRG,D-DMRG)~\cite{[held]}.  The relevant DMFT formalism 
has been developed and used with very good success~\cite{[laadv2o3]} for
a variety of problems, and so we do not reproduce it here. The only input to the DMFT is
the ``free'' density-of-states, given by $\rho_{0}^{a}(\epsilon)=\sum_{k}\delta(\omega-\epsilon_{a}(k))$.  We restrict ourselves to the quantum paramagnetic 
phase, above the $3D$ Neel ordering temperature, $T_{N}$.  With a non-local
hybridization, the Green function (GF) 
is a $2\times 2$ matrix, $G_{ab}({\bf k},\omega)$, (with $a,b=1,2$) in orbital space, as are the {\it local} self-energies,
$\Sigma_{ab}({\bf k},\omega)=\Sigma_{ab}(\omega)$.  These equations are similar
to those appearing in the DMFT for the EALM~\cite{[georges-rev]}.  The diagonal
GFs (and self-energies) yield the total {\it many body} spectral function,
$A({\bf k},\omega)=(-1/\pi)$Im[$1/(\omega-\Sigma_{a}(\omega)-\epsilon_{a}({\bf k})-t_{m}^{2}({\bf k})G_{b}({\bf k},\omega))$].
The local DOS is just $\rho_{a}(\omega)=\sum_{\bf k}A_{a}({\bf k},\omega)$.
  Using the spectral theorem, the off-diagonal spectral function,
$\rho_{12}({\bf k},\omega)=(-1/\pi)$Im$G_{12}({\bf k},\omega)$, is easily seen 
to describe $d$-wave particle-hole (excitonic) order (cf. $d$-wave form of
$t_{m}({\bf k})$).  The CT-MI (found below with QC+DMFT) thus has $d$-wave p-h
order, shown by the fact that $\langle c_{1\sigma}^{\dag}c_{2\sigma}\rangle=(-1/\pi)\sum_{\bf k}\int$ Im$G_{12}({\bf k},\omega)d\omega=0$, as shown in the lower
 inset of Fig~\ref{fig1} .  Our two-band Hamiltonian will also yield circulating current (CC) order
at finite doping concentration, $x$, as proposed by Varma~\cite{[cmv]}.  
Here, however, the {\it apical} $p-d$ link is crucial for CC order with finite 
$\Delta_{12}=i\langle (c_{1i\sigma}^{\dag}c_{2,i+e_{x,y},\sigma}-h.c)\rangle =\langle T_{i}^{y}\rangle \ne 0$ (see below).  This can be readily
seen in the large-$U$ limit of our model, where a second-order in $t/U$ 
expansion gives the ``exchange'' part as~\cite{[cmv],[dag]}:

\be
H=\frac{2}{U'-J}\sum_{<i,j>}P_{i,j}(T_{i,j}^{(1)}+T_{i,j}^{(2)})
\ee

 with $P_{i,j}=({\bf S}_{i}.{\bf S}_{j}-1/4)$, while
$T_{i,j}^{(1)}=(t_{aa}^{2}+t_{bb}^{2}+2t_{ab}^{2})(T_{i}^{z}T_{j}^{z}-1/4)+t_{aa}t_{bb}(T_{i}^{+}T_{j}^{-}+h.c)$
and
$T_{i,j}^{(2)}=t_{ab}^{2}(T_{i}^{+}T_{j}^{+}+T_{i}^{-}T_{j}^{-})+(t_{ab}(t_{aa}-t_{bb}))(T_{i}^{z}T_{j}^{x}+T_{i}^{x}T_{j}^{z})$, where 
$T_{i}^{z}=(n_{1i}-n_{2i})/2, T_{i}^{+}=c_{1i}^{\dag}c_{2i}, T_{i}^{-}=c_{2i}^{\dag}c_{1i}$.  At mean-field level~\cite{[cmv]}, (notice the difference in our
$T^{a}$s) this will yield a finite CC order, $\langle T_{i}^{y}\rangle\ne 0$~\cite{[cmv]}.
 However, our CC pattern involves {\it both} planar and apical states, i.e., it 
is closer to that proposed by Weber {\it et al.}~\cite{[circ]}, involving
three oxygens on faces of the octahedra.  While this supports the 
view~\cite{[cmv]} that $p-d$ interactions are important in cuprates, our CC 
pattern is different in details. 

  We now present our results.  In Fig~\ref{fig1}, we show the ``unperturbed''
DOS (using QC) for our two-orbital model as dash-line curves.  Clearly, the 
planar states dominate at $E_{F}$, which lies very close to a van-Hove singularity (vHs); the apical states lie much ($\simeq 1.0$~eV) lower.  The many-body spectra within QC+DMFT, shown in bold lines, are dramatically different.  Clear
lower- and upper Hubbard bands are visible in the DOS for both orbitals.  
Dramatic and large-scale spectral weight transfer (SWT), driven by strong
$U,U'$, is readily manifest.  The CT-Mott gap equals $\Delta_{MH}=1.1$~eV.  
Our result can now be directly compared with the experimental 
photoemission (PES)~\cite{[pes]} results for $La_{2}CuO_{4}$ above $T_{N}$.  

\begin{figure}[thb]
\includegraphics[width=\columnwidth]{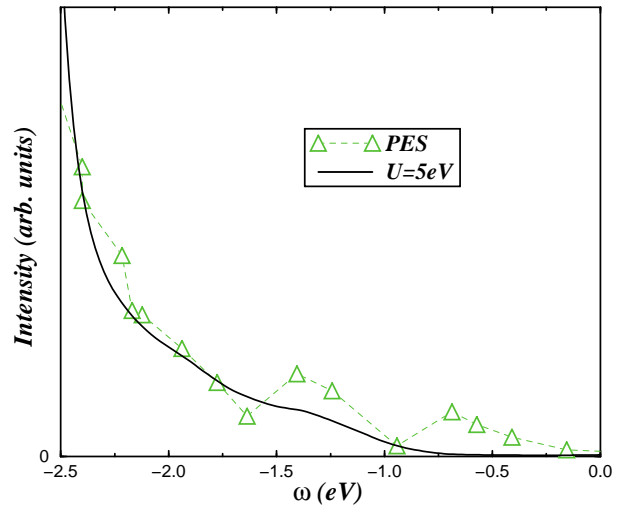}
\caption{(Color Online) Comparison of the total one-hole spectral function
obtained experimentally~\cite{[pes]} (triangles) with the QC+DMFT spectrum
with $U=5.0$~eV (full line).
Very good quantitative agreement is clearly seen up to binding energy $\simeq 2.5
$~eV.}
\label{fig2}
\end{figure}

In Fig~\ref{fig2}, we show this comparison.  Quite remarkably, very good 
quantitative agreement with the PES spectrum upto $-2.5$~eV is clearly visible.
The planar Zhang-Rice-like (ZR) states are the major contributor at low energy, but the apical
states also contribute noticeably for energies $\omega\ge 1.3$~eV.  Further,
from the self-energy (not shown), we estimate an effective mass enhancement
of $m^{*}/m_{b}\simeq 4.0$, where $m_{b}$ is the LDA band mass.  As a result, 
we get $(t,t',t'')\simeq (0.14, 0.01, 0.075)$~eV, and the resulting 
{\it renormalized} dispersion quantitatively fits the dispersion of the 
one-hole (``ZR-like'') states in ARPES~\cite{[liviu-mukul]}  {\it and} the FS 
evolution with $x$, as $E_{F}$ shifts downward
with hole doping.  We show below that our QC+DMFT yields good
quantitative agreement with the optical conductivity as well.   

  The theoretical optical conductivity is compared with the experimental data
for $La_{2}CuO_{4}$ {\it above} $T_{N}$.  In our two-band model, $\sigma(\omega)$ has two contributions: intraband transitions involving the planar- and apical states, and {\it inter}-band contributions involving transitions betwen the two
bands.  The usual DMFT equation for $\sigma(\omega)$ now reflects both these 
processes~\cite{[baldassare]}, and reads

\be
\sigma(\omega)=\sigma_{0}\int d\epsilon \rho_{0}(\epsilon) 
\int d\nu \frac{f(\omega+\nu)-f(\nu)}{\omega}\rho_{\epsilon}(\omega+\nu)\rho_{\epsilon}(\nu)
\ee
where $\rho_{\epsilon}(\omega)=\sum_{a=1}^{2} \rho_{\epsilon_{a}}(\omega)=(-1/\pi)\sum_{a=1}^{2} Im[1/(\omega+\mu_{a}-\epsilon_{a}-\Sigma_{a}(\omega))]$.
  
\begin{figure}[thb]
\includegraphics[width=\columnwidth]{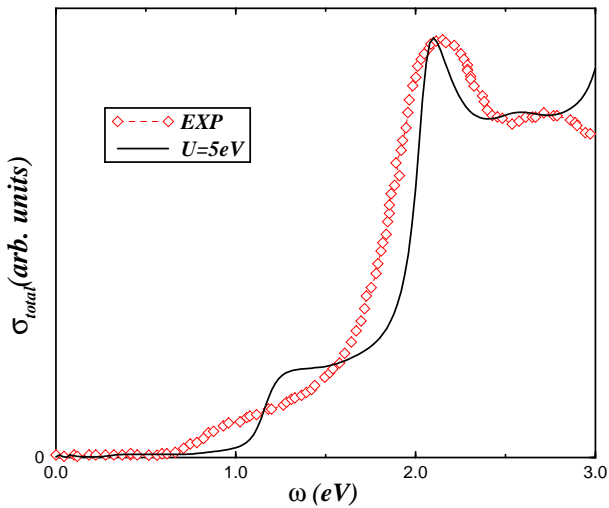}
\caption{(Color Online) Comparison of the experimental optical conductivity
(diamonds), taken from~\cite{[moore]} with the QC+DMFT result (solid line) for
$La_{2}CuO_{4}$.  Very good quantitative agreement, including description
of the $\simeq 1.0$~eV feature {\it below} the charge-transfer-Mott gap, is
clearly seen.
}
\label{fig3}
\end{figure}

  Very good quantitative theory-experiment comparison is clearly seen in
Fig~\ref{fig3}.  Specifically, the relatively sharp peak-like structure at
$\Omega=2.0$~eV, as well as the weaker shoulder (at $\Omega_{s}\simeq 1.2$~eV)
and the high-energy bump (at $\Omega_{b}\simeq 2.4$~eV) are all in satisfying 
agreement with experiment~\cite{[moore]}.  
Especially interesting is the shoulder-like feature
at $\Omega_{s}$: given that $\Delta_{MH}=1.1$~eV, one would associate the 
$2.0$~eV structure with the onset of the corresponding optical absorption
feature.  The shoulder at $\Omega_{s}$ must then be interpreted in terms of  
 a quasi-continuum ``excitonic'' feature pulled down below the Mott gap.  In our
 two-band model, this arises directly from the {\it inter}-orbital transitions 
involving the planar- and apical bands.  Thus, remarkably, QC+DMFT achieves 
very good {\it quantitative} agreement with {\it both}, one- and two-particle 
dynamical responses in the CT-MI phase, benchmarking its efficacy.

What do we expect at finite doping, $x$?
For small $x$, we expect the lower-lying apical ($d_{z^{2}}-p_{z}$) band to 
remain Mott-localized, as is generic in MO-Hubbard models~\cite{[laadv2o3],[kotliar]}, while the planar ZR band will selectively metallize.
This would then be interpretable as an ``orbital selective'' Mott transition 
(OSMT).  One should then expect {\it nodal}
fermionic QPs to dominate the responses at small $x$~\cite{[kotliar]}, also found in earlier QC work~\cite{[liviu-mukul]}.  
This OSMT would thus  
realize the famed N-AN dichotomy~\cite{[arpes],[raman]} ubiquitous to 
cuprates.  Eventually, around a critical doping, we expect the apical $p-d$ 
band to metallize as well, as $E_{F}$ shifts progressively downward with $x$.
It is tempting to link the OSMT, where strong scattering between Mott 
localized (apical) and quasi-itinerant (planar) band carriers within DMFT would lead to low-energy infra-red singularities via the Anderson {\it orthogonality 
catastrophe}, to the doping 
driven  avoided (pre-empted by $d$-SC) ``quantum critical point'' (QCP) around 
optimal doping, where low-energy singularities indeed dominate the ``strange 
metal''~\cite{[pwa],[varma]}.  More work is needed to check this theoretically, and to see whether this coincides with a $T=0$ melting of CC order~\cite{[cmv]}.
  These issues will be addressed in detail in forthcoming work.   

  In conclusion, we have proposed a new QC+DMFT method to compute the correlated
 electronic structure, along with the one- and two-particle spectra, for Mott
(CT-Mott) insulators.  Extending our earlier QC results, where very good 
agreement with the {\it dispersion} of one-hole ``ZR-like'' states {\it and} 
FS was found, we have shown how marrying QC with DMFT 
shows that, at ``high'' $T>T_{N}$, the CT-Mott insulator has $d$-wave order.
The very good {\it quantitative} agreement with {\it both} PES and optical
conductivity spectra in the
insulating state of $La_{2}CuO_{4}$ constitutes strong support for this 
conclusion.  Our QC+DMFT modelling thus
reconciles the ARPES and dHvA results with one- and two-particle spectral
responses for the CT-MI phase.  
In light of recent ideas~\cite{[arpes],[raman],[pwa],[cmv]}, these 
findings serve as an excellent starting point to study the physics of doped 
cuprates in 
detail within an {\it ab-initio} correlated approach.  QC+DMFT should also 
serve as a new theoretical tool with wide application to other correlated 
systems of great current interest.

{\bf Acknowledgements} We thank Prof. P. Fulde for useful discussions.

\end{document}